\documentstyle[12pt]{article}

\newcounter{popnr}

\def\sec{\setcounter{equation}{0}}

\def\be{\begin{equation}}
\def\ee{\end{equation}}

\begin{document}

\baselineskip20pt

\title{Gauging of 1d-space translations\\ for nonrelativistic matter -- geometric bags}

\author{P.C. Stichel\\An der Krebskuhle 21, D-33619 Bielefeld\thanks{email: hanne@physik.uni-bielefeld.de}}

\date{\today}

\maketitle

\begin{abstract}
We develop in a systematic fashion the idea of gauging $1d$-space translations with fixed Newtonian time for nonrelativistic matter (particles and fields). By starting with a nonrelativistic free theory we obtain its minimal gauge invariant extension by introducing two gauge fields with a Maxwellian self interaction. We fix the gauge so that the residual symmetry group is the Galilei group and construct a representation of the extended Galilei algebra. The reduced $N$-particle Lagrangian describes geodesic motion in a $(N-1)$-dimensional (Pseudo-) Riemannian space. The singularity of the metric for negative gauge coupling leads in classical dynamics to the formation of geometric bags in the case of two or three particles. The ordering problem within the quantization scheme for $N$-particles is solved by canonical quantization of a pseudoclassical Schr\"odinger theory obtained by adding to the continuum generalization of the point-particle Lagrangian an appropriate quantum correction. We solve the two-particle bound state problem for both signs of the gauge coupling. At the end we speculate on the possible physical relevance of the new interaction induced by the gauge fields.
\end{abstract}

\newpage

\sec
\section{Introduction}

Gauging a symmetry may be called the leading principle for the construction of fundamental interactions in physics (cp. [1]).

Not only internal symmetry groups are gauged ($U(1)$ in QED, $SU(N)$ in QCD) but space time symmetries are gauged too (Poincar\'e group in general relativity).

But the gauge principle not only determines a vast amount of relativistic physics as elementary or gravitational physics, it is also of importance in condensed matter physics (description of the Fractional Quantum Hall effect by an abelian gauge field with a Chern-Simons term minimally coupled to charged matter). Therefore we may ask the question of the importance of the gauge principle in nonrelativistic physics quite generally. What about gauging the Galilei group? In recent work by De Pietri et al. [2] this task has been taken up for a single point particle and extended systems in $(3+1)$-dimensions. But the authors of [2] started with the nonrelativistic limit of general relativity and threw away all fields not coupled to matter in this limit. In this respect our work will be completely different from [2]. Our leading principles are:

\begin{description}
\item{i)} We begin with a nonrelativistic free theory of matter which is invariant with respect to global Galilei transformations,
\item{ii)} we ask for the smallest number of gauge fields with a minimal coupling Lagrangian leading to a gauge invariant theory.
\end{description}

In contrast to [2] the main aim of our gauging procedure is the determination of the interaction between the constitutents of matter (point particles or fields) induced by the gauge fields.

In a recent letter [3] we started to work out this idea with the simplest example: Classical point particles moving on a line or on a circle. 
The underlying global transformations are $1d$-space translations
whose gauging leads to general coordinate transformations at fixed
Newtonian time. The minimal coupling Lagrangian for this example contains only two gauge fields with a Maxwellian interaction term. 
In particular we don't introduce an
additional dilaton field as in $(1+1)$-gravity [4]. The arbitrary
gauge function in the solution for the gauge field has been chosen such,
that the Galilei group appears as the residual symmetry group.
In the present paper we will give a full account of [3] for point particles on the line, extend the work to the case of fields and treat the quantization of particles and fields.

Our paper is organized as follows: In section 2 we review the point particle case from [3] and give a continuum generalization in terms of hydrodynamic field variables. In addition we discuss the local conservation law following from Noether's theorem. Section 3 is devoted to a full discussion of the gauge fixing problem with Galilei symmetry as the resulting residual symmetry. In section 4A we give the reduction of our Lagrangian to matter variables by using the symplectic Hamiltonian procedure and discuss the resulting $N$-particle dynamics as geodesic motion in a $(N-1)$--dimensional (Pseudo-) Riemannian space. The representation of the extended Galilei algebra with the total mass as central charge is treated in section 4B. We continue the discussion of classical physics with the string-like behaviour of the two-particle system in section 4C. In particular we obtain a dynamically determined geometric bag model for two or three particles in the case of a negative gauge coupling (section 4D). After a discussion of the ordering problem within the quantization scheme for point particles in section 5A we enlarge our hydrodynamic field theory to a pseudo-classical Schr\"odinger theory by adding a quantum correction term (section 5B). The canonical quantization of this Schr\"odinger theory will be treated in section 5C. In this way we give a solution of the ordering problem for the N-particle Hamilton operator. Section 6 gives an analytic solution for the two-particle bound state problem for both signs of the gauge coupling. Finally, section 7 contains some final remarks including an outlook on subsequent work.

\sec
\section{Minimal-coupling Lagrangian}
\subsection*{A) Point particles}

We start with $N$ nonrelativistic particles in free motion on a line
$({\bf R}^1)$ described by the Lagrangian\footnote{For reasons of simplicity we give all particles the same mass $m=1$ in appropriate units.}
\be
L_0 = {1 \over 2} \sum^N_{\alpha = 1} (\dot{x}_\alpha (t))^2
\ee
The equations of motion (EOM) following from (2.1)
\be
\ddot{x}_\alpha = 0
\ee
are invariant with respect to global Galilei-transformations
\be
(x,t) \to (x^\prime, t^\prime)
\ee
with
\be
x^\prime = x + a + vt
\ee
and
\be
t^\prime = t + b
\ee
where the parameters $a, v$ and $b$ take values in ${\bf R}^1$.
Now we generalize (2.4) to a local transformation, given in infinitesimal form by
\be
\delta x = a (x,t)
\ee
where $a(x,t)$ is an arbitrary, twice differentiable and bounded function of its arguments.
Eq. (2.6) describes local space translations (including local
boosts). We keep time translations with a constant $b$ (cp. (2.5)).

Obviously the EOM (2.2) are not invariant with respect to the transformation 
(2.6). In order to repair that, we introduce two gauge fields $h (x,t)$ and
$e (x,t)$ and replace $\dot{x}$ for each particle in (2.1) by the function\footnote{Our procedure differs from the corresponding one in [2] applied to one space dimension. In [2] $L_0$ would be replaced by a polynomial of second order in $\dot{x}$ requiring three gauge fields instead of two.} $\xi$
\be
\xi = h (x,t) \dot{x} + e (x,t).
\ee
Invariance of $\xi$ with respect to (2.6) requires the following transformation
rules for the gauge fields
\be
\delta h = - h \partial_x a~,~\delta e = - h \partial_t a
\ee

\noindent
where 
\be
\delta f (x,t): = f^\prime (x + \delta x, t + \delta t).
\ee
With (2.7) we supply the minimal gauge invariant extension for the particle velocities 
$\dot{x}_\alpha$.

In agreement with $(1+1)$-gravity (cp. [5]) we conclude from (2.7) and (2.8) that our gauge fields $e(x,t)$ and $h(x,t)$ are components of the Zweibein $\{ E^b_\mu\}$ with components $\{ E^0_\mu \} ) = (1,0)$ and $\{ E^1_\mu \} = (e,h)$. In our case the vector $\{ E^0_\mu\}$ is trivial because time has been held fixed up to translations with a constant.

Now $L_0$ in (2.1) has to be replaced by
\be
L_{\rm matter} = {1 \over 2} \sum^N_{\alpha = 1} (\xi_\alpha (t))^2
\ee
with
\be
\xi_\alpha (t): = h_\alpha (t)  \dot{x}_\alpha (t) + e_\alpha
(t)
\ee
where we defined
\be
f_\alpha (t) : = f(x_\alpha(t),t)
\ee
for an arbitrary field $f(x,t)$. 

We must supplement (2.10) by an invariant (or quasi-invariant) Lagrangian
${\rm L_{\rm field}}$ describing the self-interaction of the gauge fields.
Let us define a field strength $F$
\be
F: = {1 \over h} (\partial_t h - \partial_x e)
\ee

From (2.9) we obtain easily the commutator between $\partial$ and any partial differentiation $\partial \in (\partial_t, \partial_x)$
\be
\delta \partial f = \partial \delta f - (\partial a) \partial_x f\ .
\ee

We infer from (2.8) and (2.14) that our field strength $F$ is gauge invariant
\be 
\delta F = 0\ .
\ee
Therefore, any integral of the form
\be
\int_{{\bf R}^1} d\mu_t (x) K (F(x,t))
\ee
with the invariant measure
\be
d\mu_t (x): = h (x,t) dx
\ee
is a candidate for ${\rm L_{\rm field}}$.
The simplest, nontrivial example for $K$ is a quadratic
\be
K(Z) = Z^2
\ee
With this Maxwellian choice for $\rm L_{\rm field}$ our action takes the
form
\be
S = \int dt ({\rm L_{\rm matter}} + {\rm L_{\rm field}})
\ee
with
\be
{{\rm L}_{\rm field}} = {1 \over {2\lambda}} \int dx h (x,t) F^2 (x,t)
\ee
where $\lambda$ is a coupling strength. With our Zweibein $\{ E^b_\mu\}$ we identify $hF$ as the only nonvanishing component of the torsion tensor and (2.20) as the related local quadratic Lagrangian (cp.~[7]). 

By varying $S$ with respect to $x_\alpha$ we get the particle-EOM
\be 
\dot{\xi}_\alpha + \xi_\alpha F_\alpha = 0\ .
\ee
For the following it is advantageous to replace the Lagrangian (2.10) by its
1st-order form
\be
{{\rm L}_{\rm matter}} = \sum_\alpha \xi_\alpha (h_\alpha \dot{x}_\alpha + e_\alpha ) - \frac{1}{2} \sum_\alpha \xi^2_\alpha 
\ee
Variation of $S$ with respect to $\xi_\alpha$ leads to (2.11) now as a constraint rather than a definition.

\subsection*{B) Fields}

The continuum generalisation of the Lagrangian (2.22) will be given in analogy to the case of free particles [6] in terms of hydrodynamics variables $\rho(x,t)$ and  $\theta(x,t)$ describing a density field and a potential field respectively. Therefore we look for a Lagrangian $\rm L_{\rm matter}$ such that for the particular choice\footnote{The factor $1/h$ in front of the r.h.s. of (2.23) arises from the fact that $\rho(x,t)$ is a density with respect to the measure $d\mu_t (x)$.} 
\be
\rho (x,t) = \frac{1}{h(x,t)} \sum_\alpha \delta (x-x_\alpha (t)) 
\ee
and the identification
\be
\xi (x,t) : = \frac{1}{h(x,t)} \partial_x \theta (x,t)
\ee
we obtain up to a total time-derivative the expression (2.22).

It can easily be shown that the desired expression for $\rm L_{\rm matter}$ is given by
\begin{eqnarray}
{{\rm L}_{\rm matter}} &=& - \int d \mu_t (x) \rho (x,t) (D_t \theta)(x,t)\nonumber\\
& & - \frac{1}{2} \int d \mu_t (x) \rho (x,t) (D\theta)^2 (x,t)
\end{eqnarray}
where the invariant derivatives $D$ and $D_t$ are defined by means of the inverse Zweibein $\{ \tilde{E}^\mu_b\}$
\begin{eqnarray} 
D &: =&  \tilde{E}^1_1 \partial_x = \frac{1}{h(x,t)} \partial_x\\
{\rm and} \quad \quad D_t &: =& \tilde{E}^\mu_0 \partial_\mu = \partial_t - e (x,t) D
\end{eqnarray}

Under the proviso that our hydrodynamic field $e$ and $\theta$ are invariant with respect to local coordinate transformations (2.6)
\be
\delta \rho = \delta \theta = 0
\ee
the invariance of $\rm L_{\rm matter}$
\be
\delta \rm L_{\rm matter} = 0
\ee
follows.

\noindent
This invariance leads by means of Noether's 2nd theorem to a local conservation law 
\be \partial_t (\rho \partial_x \theta) + \partial_x (j D \theta - \frac{1}{2\lambda} F^2) = 0
\ee
where the current $j$ is defined by
\be
j : = \rho (D\theta - e)\ .
\ee
As for any gauge theory the local conservation law (2.30) is nothing but a combination of the EOM for gauge fields and defines a superpotential (cp. [8]):

\noindent
By varying our action with respect to $h$ and $e$ respectively we obtain 
\be
\partial_t F = \lambda j D \theta - \frac{1}{2} F^2
\ee
and 
\be
\partial_x F = - \lambda \rho \partial_x \theta\ .
\ee
In order that the boundary term in the derivation of (2.33) vanishes we assumed that $e$ and $h$ are finite at spatial infinity and $F$ vanishes there. The latter requirement leads due to (2.33) to the constraint
\be
\int dx \rho \partial_x \theta = 0\ .
\ee
By inserting (2.32-33) the local conservation law (2.30) is satisfied identically. It follows that the superpotential is proportional to the field strength $F$ as in electrodynamics. But the corresponding charge vanishes in our case.

Furthermore due to the invariance of $\rm L_{\rm matter}$ with respect to the transformation 
\be
\theta \to \theta +\ \mbox{const.}
\ee
we obtain the continuity equation
\be
\partial_t (h\rho) + \partial_x j = 0\ .
\ee

\sec
\section{Gauge fixing}

In this section we will show, that the gauge field $h(x,t)$ is not a dynamical variable but a kinematical entity to be determined by fixing the class of physically admissible frames of reference.

Let us start with introducing a gauge function $\Lambda$ by
\be
h(x,t) = \partial_x \Lambda (x,t)
\ee

With the local transformation properties (2.8) and (2.14) we infer from (3.1)

\noindent
\begin{description}
\item{i)} The most general form of $e(x,t)$ is given by
\begin{equation}
e(x,t) = \partial_t \Lambda (x,t) + \hat{e} (x,t)
\end{equation}
with $\delta \hat{e} = 0$ (but in general $\hat{e}$ depends on $\Lambda$!)

\item{ii)} $\delta \Lambda = 0$, or for finite transformations
generated by $a(x,t)$
\be
\Lambda^\prime (x,t) = e^{-a(x,t)\partial_x} \Lambda (x,t)
\ee
\end{description}

\noindent
Physically admissible frames of reference in the nonrelativistic regime are inertial frames connected to each other by Galilei transformations. Therefore we have to fix a class of functions $\{ \Lambda^\prime (x,t)\}$ such that the remaining (residual) symmetry group is the Galilei group only. This means we must look for  $\{ \Lambda^\prime\}$ invariant with respect to transformations generated by
\be
\tilde{a} (x,t) = \alpha + \beta t
\ee
only. This is the case iff
\be
\Lambda^\prime (x,t) = x - (a + vt)
\ee
which is equivalent to fixing the gauge by 
\be
h(x,t) = 1
\ee
and demanding
\be
\delta e = - \delta \beta\ .
\ee

\noindent
It remains to show that for an arbitrary function $\Lambda(x,t)$ with $\partial_x \Lambda \not= 0$ it is always possible to find a transformation 
\be
x \to x^\prime(x,t)
\ee
such that $\Lambda^\prime (x^\prime,t)$ is given by (3.5). Obviously this goal will be achieved with 
\be
x^\prime (x,t) = \Lambda (x,t) + (a+vt)
\ee
since (3.3) may be written as
\be
\Lambda^\prime (x^\prime,t) = \Lambda (x,t)\ .
\ee

\sec
\section{Classical dynamics}

\subsection*{A) Reduced matter Lagrangian}

In order to express our Lagrangian $L$ in terms of matter variables only,
we will follow the symplectic Hamiltonian procedure of Faddeev and Jackiw [9].

We enlarge our phase space by an additional field $\pi (x,t)$ and obtain for the canonical 1st order form of $L$
\be
L = \int dx (- \rho \partial_t \theta - {\cal H} (x,t))
\ee
with (after a partial integration)
\begin{eqnarray}
{\cal H} (x,t) : &=& \frac{1}{2} \rho (\partial_x \theta)^2 + \frac{\lambda}{2} \pi^2\nonumber\\
& & - e (\partial_x  \pi + \rho \partial_x \theta)\ .
\end{eqnarray}

In (4.2) $e(x,t)$ is a Lagrange-multiplier field. Variation of $S$ with respect to $e$ 
leads to the constraint
\be
\partial_x \pi = - \rho \partial_x \theta 
\ee
which has the solution
\be
\pi (x,t) = - \frac{1}{2} \int dy \epsilon (x-y) \rho (y,t) \partial_y \theta (y,t)
\ee
where $\epsilon (x) : = x/|x|$. 

In order that the boundary term arising in the derivation of (4.1-2) vanishes and to obtain $\mid \int dx \pi^2 \mid < \infty$ we have to require a boundary condition at spatial infinity:
\be
\pi \mid_{x=\pm\infty} = 0\ .
\ee
From (4.4) we conclude that (4.5) implies the constraint (cp. (2.34))
\be 
\int d x \rho (x,t) \partial_x \theta (x,t) = 0
\ee
i.e. the total canonical momentum of matter has to vanish.

The constraint (4.6) eq. can't be solved explicitly in terms of dynamical variables. Therefore we have to take into account (4.6) by means of a Lagrange multiplier $v$. 
With that and the insertion of (4.4) into (4.2) we obtain for the reduced matter Lagrangian ${\rm L}_{red}$
\be
{\rm L}_{red} = \int dx (- \rho \partial_t \theta - {\rm H}_{red} (x,t)
\ee
with 
\be
{\rm H}_{red} (x,t) : = \frac{1}{2} (\partial_x \theta)^2 + \frac{\lambda}{2} \pi^2 + v \rho \partial_x \theta
\ee
where $\pi (x,t)$ is given by (4.4).

By varying ${\rm L}_{red}$ with respect to $\rho$ and $\theta$ we obtain the hydrodynamic EOM
\be
\partial_t \theta = - \frac{1}{2} (\partial_x \theta)^2 + e \partial_x \theta
\ee
and the continuity equation (2.36) respectively where the gauge field $e(x,t)$ is given in terms of $\pi (x,t)$ as follows
\be
e(x,t) = - v - \frac{\lambda}{2} \int dy \epsilon (x-y) \pi (y,t)\ .
\ee
Using (4.4) and (4.6) $e(x,t)$ may be rewritten as
\be
e(x,t) = - v + \frac{\lambda}{2} \int dy |x-y| \rho (y,t) \partial_y \theta (y,t)\ .
\ee

The Lagrange multiplier $v$ will not be determined from the EOM. 
In accordance with section 3 we choose $v$ as constant, i.e. time independent.

\medskip
Comparison of the EOM's derived from ${\rm L}_{red}$ with the correponding Poisson-bracket (PB) relations $\dot{A} = \left\{ A, {\rm H}_{red}\right\}$ for any field $A$ leads to the symplectic structure
\be
\left\{ \rho (x,t) , \theta (y,t) \right\} = \delta (x-y)
\ee
with all other PB's vanishing.

We note that ${\rm L}_{red}$ is invariant with respect to time translation too. Therefore our residual symmetry is given by the full Galilei group.

By means of (2.23-24) we may specialize (4.7) to the case of $N$ point particles. We obtain after a partial integration and use of (4.6) up to a total time derivative
\be
{{\rm L}_{red}} = \sum_\alpha \xi_\alpha (\dot{x}_\alpha - \frac{1}{2} \xi_\alpha - v) + \frac{\lambda}{4} \sum_{\alpha,\beta} |x_{\alpha\beta}| \xi_\alpha \xi_\beta
\ee
where $x_{\alpha\beta} : = x_\alpha - x_\beta$. 

\medskip
\noindent
From (4.6) we read off the constraint 
\be
\sum^N_{\alpha =1} \xi_a = 0
\ee
which may be solved for $\xi_1$ leading to (using the summation convention, indices running from 2 to $N$)
\be
{\rm L}^\prime_{red} = \xi_\alpha \dot{\eta}^\alpha - {{\rm H}_{red}}
\ee
with
\be
{\rm H}^\prime_{red} = \frac{1}{2} g^{\alpha\beta} (\eta) \xi_\alpha \xi_\beta
\ee
where
\[
\eta^\alpha : = x^\alpha - x^1
\]
and
\be
g^{\alpha\beta}(\eta) : = \delta^{\alpha\beta} + 1 + \frac{\lambda}{2} \left( |\eta_\alpha | + |\eta_\beta | - |\eta_\alpha - \eta_\beta|\right)
\ee

\noindent
${\rm L}_{red}^\prime$ is now independent of $v$. Galilei symmetry becomes a hidden symmetry.

\medskip
\noindent
We observe that ${\rm H}_{red}^\prime$ looks like a free Hamiltonian describing the geodesic motion of one-'particle' in a $(N-1)$-dimensional (Pseudo-) Riemannian space $M$ with metric $\{ g_{\alpha\beta}\}$ (obtained as the inverse of (4.17)).

By varying the action $S^\prime = \int dt {\rm L}_{red}^\prime$ with respect to $\xi_\alpha$ we obtain the constraint
\be
\dot{\eta}^\alpha = g^{\alpha\beta} (\eta) \xi_\beta\ .
\ee
If we express ${\rm L}_{red}^\prime$ and ${\rm H}_{red}^\prime$ by means of (4.18) in terms of $(\{ \eta^\alpha\}, \{ \dot{\eta}^\alpha\})$ we obtain
\be
{\rm L}_{red}^\prime = {\rm H}_{red}^\prime = \frac{1}{2} g_{\alpha\beta} (\eta) \dot{\eta}^\alpha \dot{\eta}^\beta
\ee
Therefore the Euler-Lagrange equations become the geodesic equations in standard form
\be
\ddot{\eta}^\alpha + \Gamma^\alpha_{\beta\gamma} \dot{\eta}^\beta \dot{\eta}^\gamma = 0
\ee
where the $\Gamma^\alpha_{\beta\gamma}$ are the Christoffel-symbols
\[
\Gamma^\alpha_{\beta\gamma} : = \frac{1}{2} g^{\alpha\delta} (\partial_\beta (g_{\gamma\delta}) + \partial_\gamma (g_{\delta\beta}) - \partial_\delta (g_{\beta\gamma}))\ .
\]
On the other hand the variations of $S^\prime$ with respect to $\eta^\alpha$ leads to the EOM
\be
\dot{\xi}_\alpha = - \frac{1}{2} \partial_\alpha (g^{\beta\gamma}) \xi_\beta \xi_\gamma\ .
\ee 
It is self-evident that the combination of (4.18) and (4.21) leads back to (4.20). 

\medskip
Writing (4.18) and (4.21) in PB-form we infer from (4.16) the canonical symplectic structure
\be
\{ \eta^\alpha, \xi_\beta\} = \delta^\alpha_\beta\ , \qquad \{ \eta^\alpha, \eta^\beta\} = \{ \xi_\alpha, \xi_\beta\} = 0
\ee

\subsection*{B) Representation of the extended Galilei algebra}

We have pointed out in section 3 that with the choosen gauge (3.6-7) the residual symmetry group is the Galilei group.

In this section we will show that our classical theory described in section 4A allows a representation of the extended Galilei algebra with one central charge given by the total mass of the system.

In order to have the correct physical insight into each step of the following treatment we consider the case of $N$ point particles first. Afterwards we generalize the results for the case when we are dealing with fields.

It is well known that the appearance of a central charge is related to the quasi invariance of the Lagrangian with respect to the symmetry transformation considered. But our ${\rm L}_{red}$ (eq. (4.13)) is invariant with respect to translations and boosts. In order to change this situation  for boosts we add to ${\rm L}_{red}$ a total time derivative (which leaves the EOM unchanged)
\be
\frac{d}{dt} (v \sum_\alpha (x_\alpha - \frac{v}{2} t))
\ee
and obtain the 1st order Lagrangian
\be
\tilde{\rm L}_{red} = \sum_\alpha \dot{x}_\alpha (\xi_\alpha + v) + \dot{v} \sum_\alpha (x_\alpha -vt) - \tilde{{\rm H}}_{red}
\ee
with 
\be
\tilde{{\rm H}}_{red} : = \frac{1}{2} \sum_\alpha (\xi_\alpha^2 + v^2) + v\sum_\alpha \xi_\alpha - \frac{\lambda}{4} \sum_{\alpha,\beta} |x_{\alpha\beta} | \xi_\alpha \xi_\beta\ .
\ee
Now $ \tilde{{\rm L}}_{red}$ is quasiinvariant with respect to infinitesimal boosts $\delta x_\alpha = t\delta\beta$, $\delta v = \delta \beta$
\be
\delta  \tilde{{\rm L}}_{red} = \delta \beta \frac{d}{dt} \sum_\alpha x_\alpha\ .
\ee
From (4.24-25) we infer, that our extended phase space will be described by the set of variables $(\{ x_\alpha\}, \{p_\alpha\}, v, p_v)$ where
\begin{eqnarray}
p_\alpha : = \frac{\partial  \tilde{{\rm L}}_{red}}{\partial\dot{x}_\alpha} &=& \xi_\alpha + v\\
p_v : = \frac{\partial  \tilde{{\rm L}}_{red}}{\partial \dot{v}} &=& \sum_\alpha (x_\alpha - vt)\ .
\end{eqnarray}
We remark that (4.28) is a constraint reducing the degrees of freedom in phase space, but for the following it is necessary to stay in extended phase space.

We define the total canonical momentum $P$ by
\be
P : = \sum_\alpha p_\alpha
\ee
and observe that it has the correct physical behaviour with respect to boosts
\be
\delta P = N\delta\beta\ .
\ee
$ \tilde{{\rm H}}_{red}$ expressed in terms of phase space variables reads
\be
\tilde{{\rm H}}_{red} = \frac{1}{2} \sum_\alpha p_\alpha^2 - \frac{\lambda}{4} \sum_{\alpha,\beta} |x_{\alpha\beta}| (p_\alpha -v)(p_\beta -v)\ .
\ee 

From (4.29) and (4.31) we obtain 
\be
\{ P ,  \tilde{{\rm H}}_{red}\} = 0\ .
\ee
By using the canonical symplectic structure for our phase space variables it is easily seen that the conserved boost generator $K$ leading to the desired PB-relations
\be
\{ K, P\} = N
\ee
and
\be
\{ K,  \tilde{{\rm H}}_{red} \} = P
\ee
is given by
\be
K = \sum_\alpha x_\alpha - p_v - tP\ .
\ee
The r.h.s. of (4.33) describes the total mass of the $N$-particle system in units of $m=1$. Therefore the equations (4.32-34) constitute a representation of the extended Galilei algebra with one central charge. 

According to (2.23-24)$_{h=1}$ the generalization of the foregoing results to the case of fields is obtained by means of the following substitutions

\medskip
\begin{description}
\item{i)} add to ${\rm L}_{red}$ (eq. (4.7)) the expression 
$$
\frac{d}{dt}\int dx \rho (\theta + vx - \frac{v^2}{2} t)
$$

\item{ii)} define total momentum $P$ and boost generator $K$ respectively by
$$
P : = \int dx \rho (\partial_x \theta + v)
$$
and
$$
K : = \int dx x \rho - p_v - tP\ .
$$
\end{description}

\subsection*{C) String-like behaviour}

In the point particle case the $\pi$-field (eq. (4.4) takes the form
\be
\pi (x,t) = - \frac{1}{2} \sum^N_{\alpha = 1} \epsilon (x-x_\alpha (t)) \xi_\alpha (t)
\ee
with the constraint (4.14). Therefore $\pi(x,t)$ vanishes outside a closed 
$x$-intervall
\be
\pi (x,t) = 0 \qquad \forall x \notin \left[ \mu_1 (t), \mu_2 (t)\right]
\ee
with $\mu_{{1\atop 2}} (t) : = {\min \atop \max} \{ x_\alpha\}$.

\medskip
\noindent
In particular for $N=2\ $ $\pi(x,t)$ is string-like with point masses at its ends.

From (4.17,20) we obtain the following EOM for relative particle motion  $(x:=x_1 - x_2)$
\be
\ddot{x} - \lambda E \epsilon (x) = 0
\ee
where the conserved energy $E$ is given by
\be
E : = \frac{1}{4}\ \frac{\dot{x}^2}{1+\frac{\lambda}{2} |x|}
\ee

We conclude

\smallskip
\noindent
\begin{description}
\item{i)} For $E > 0$ but $\lambda < 0$ the 'string' is always of finite length (bounded motion)                   

\item{ii)} The interparticle potential is proportional to their distance with strength proportional to the energy. This leads according to (4.38-39) to a constant positive (negative) acceleration for $x > 0$ $(x<0)$ for either sign of $\lambda$ (for $\lambda < 0$ if $|x| > 2/|\lambda |$ only) resulting in arbitrary large velocities if time goes on. Therefore the model is unphysical for $\lambda > 0$ and for $\lambda < 0$ if $|x| > 2/| \lambda |$.

\item{iii)} For $N=2$, $M$ is a one-dimensional Riemannian space and therefore flat. This allows by means of the transformation
\be
x \to y : = \frac{4}{|\lambda|} (1 + \frac{\lambda}{2} |x|)^{1/2}
\ee
the transition to a $1d$-Euclidean space $E_1$
\[
 E_1 = \left\{ \begin{array}{l l l}
\left[ 4/\lambda,\infty\right] & \mbox{for} & \lambda > 0\\
\left[ 0, 4/|\lambda| \right] & \mbox{for} & \lambda < 0
\end{array} \right.
\]
On $E_1$ we have the free EOM
\be
\ddot{y} = 0
\ee
where the finite end points of $E_1$ are points of reflection.

\item{iv)} For $\lambda < 0$ the metric is singular at $|x| = 2/|\lambda|$. The region $|x| > 2/|\lambda|$ is 
physically disconnected from the region $|x| < 2/|\lambda|$ (energy conservation!)
\end{description}

\subsection*{D) Geometric bags}

Let us consider the case of $\lambda < 0$ for $N$-point particles in more detail. In the last section we discovered for $N=2$ exclusively bounded motion within the region $|x| < 2/|\lambda|$ determined by the singularity of the metric at $|x| = 2/|\lambda|$. This is a geometric bag as a dynamical consequence of our gauge theory. What about $N \ge 3$~? For $N > 3$ we are unable to make any statement at present. But in the following we will show that for $N = 3$ a geometric bag arises again determined by the singularity of the metric: 

\smallskip
\noindent
Let us consider the energy $E$ for relative motion (4.19). Because of the symmetry of $E$ with respect to arbitrary permutations of the particle numbers 1, 2 and 3 it is sufficient for the following to consider the particular ordering
\be
x_1 < x_2 < x_3\ .
\ee
Then we obtain from (4.16) and (4.17) with the substitutions
\[
\eta^{i+1} \to \kappa_i : = | \lambda | \eta^{i+1} \qquad \mbox{and} \qquad
\xi_{i+1} \to \xi_i\ , \ \ \ i = 1,2
\]
\begin{eqnarray}
E &=& \frac{1}{2\lambda^2 D} ((2 - \kappa_2) \dot{\kappa}_1^2 + (2 - \kappa_1) \dot{\kappa}_2^2 - \nonumber\\
& & - 2 (1 - \kappa_1) \dot{\kappa}_1 \dot{\kappa}_2)
\end{eqnarray}
with
\be
D : = \det (g^{\alpha\beta}) = 3 - 2 \kappa_2 + \kappa_1 (\kappa_2 - \kappa_1)
\ee
where due to (4.42)
\be
0 < \kappa_1 < \kappa_2 \ .
\ee
From (4.44-45) we infer that 
\be
D \ge 0 \qquad \mbox{if} \qquad \kappa_1 < \kappa_2 \le \kappa_1 + \frac{2\kappa_1 -3}{\kappa_1 - 2}\ , \ \ \kappa_1 < \frac{3}{2}
\ee
and $D < 0$ in the adjoining region.

Therefore the two curves $C_i$, $i=1,2$
\begin{eqnarray*}
C_1 :&& \kappa_2 = \kappa_1 + \frac{2\kappa_1 -3}{\kappa_1 - 2} \qquad \mbox{with} \ \ \ 0 \le \kappa_1 \le 3/2\\
C_2 : && \kappa_2 = \kappa_1 \ge 0
\end{eqnarray*}
form the boundary of a finite region ${\bf B}_1$ in the $(\kappa_1, \kappa_2)$-plane with vanishing $D$ on $C_1$ corresponding to the singularity of our metric.

From (4.46) we conclude that $E$ is positive within ${\bf B}_1$
\be
E > \frac{1}{2\lambda^2 D (2 - \kappa_1)} ((1 - \kappa_1) \dot{\kappa}_1 - (2 - \kappa_1) \dot{\kappa}_2 )^2  > 0 \ .
\ee
But $E$ becomes infinite on $C_1$ for generic values of $\dot{\kappa}_{1,2}$. Therefore at finite energy particles can't reach the boundary $C_1$: they are confined within ${\bf B}_1$.

If we complete the picture by considering in addition to (4.42) the other five possible orderings, we arrive at a finite region ${\bf B}$ defining a geometric bag for our three particles. 

\sec
\section{Quantization}

\subsection*{A) The ordering problem}

In quantizing the $N$-particle system described by the Hamiltonian (4.16) and the symplectic structure (4.22) we have according to the canonical quantization recipe to substitute
\be
\xi_\alpha \to p_\alpha  (= \frac{\hbar}{i} \frac{\partial}{\partial\eta_\alpha} = : \frac{\hbar}{i} \partial_\alpha)
\ee
and
\be
\{ A,B\} \to \frac{\left[ A,B\right]}{i\hbar}
\ee
where [A,B] is the commutator of the operators $A$ and $B$ which denote any pair from the phase space set $\{ \eta_2, \ldots, \eta_N, p_2, \ldots, p_N\}$. But (5.1) does not specify uniquely the quantum analogon of the Hamiltonian $g^{\alpha\beta} (\eta) \xi_\alpha \xi_\beta$, because the space variables $\{ \eta^i\}$ and the momentum variables $\{ p_i\}$ become non-commuting operators. This is the well known ordering problem. The only restriction we have is the requirement of hermiticity for the Hamilton operator. But this leaves a whole family of admissible Hamilton operators for a given classical Hamiltonian. Let us illustrate this for the simplest case, $N=2$. From (4.16) we have
\be
{{\rm H}_{\rm class}} = (1 + \frac{\lambda}{2} |x|) \xi^2
\ee

\noindent
From that we obtain two fundamental hermitian Hamilton operators (with respect to the usual inner product)
\begin{eqnarray}
H_0 &=& p (1 + \frac{\lambda}{2} |x|)p\\
H_1 &=& \frac{1}{2} (p^2 (1+\frac{\lambda}{2} |x|) + (1 + \frac{\lambda}{2}|x|)p^2)\ .
\end{eqnarray}

\noindent
The most general hermitian Hamilton operator is given by a linear combination
\be
H_\gamma = \gamma H_1 + (1-\gamma)H_0\ , \qquad \gamma \in \left[ 0,1\right]\ .
\ee

\noindent
How do they differ from each other? Consider the difference $H_\gamma - H_{\gamma^\prime}$. By means of Heisenberg's commutation relations we obtain 
\be
H_\gamma - H_{\gamma^\prime} = (\gamma - \gamma^\prime) \frac{\lambda \hbar^2}{2} \delta(x)\ .
\ee

Therefore the $H_\gamma$ differ in the strength of a contact potential. 
This term beeing proportional to $\hbar^2$ is a quantum correction.

\bigskip
\noindent
But what we want to quantize are not separate theories for  different particle numbers $N$ but a appropriate field theory with a particle interpretation after quantization. This quantum field theory should lead to a $N$-particle amplitude with a definite ordering prescription for the quantum analogon of $g^{\alpha\beta} (y) \xi_\alpha \xi_\beta$. For this the hydrodynamic field theory described in section 4A is not a suitable starting point. It lacks a particle interpretation after quantization.

\subsection*{B) The pseudo-classical Schr\"odinger field}

In order to obtain a particle interpretation after quantization we have to start with a classical complex-valued field $\psi(x,t)$. 

We know of only one complex-valued field which after quantization possesses a particle interpretation in nonrelativistic physics: the Schr\"odinger field $\psi(x,t)$.

\noindent
With the Madelung representation [10]
\be
\psi(x,t) = \sqrt{\rho(x,t)} e^{\frac{i}{\hbar}\theta(x,t)}
\ee
our hydrodynamic field theory (section 4A)) becomes a Schr\"odinger field theory by adding to the Lagrangian density in (4.7) a quantum correction term $0(\hbar^2)$
\be
- \frac{\hbar^2}{8 \rho (x,t)} (\partial_x \rho (x,t))^2
\ee
The resulting Lagrangian ${\rm L}_{red}$ may be expressed in terms of $\psi$ and $\psi^\dagger$ as follows
\be
{{\rm L}_{red}} = \int dx \left[ \frac{i\hbar}{2} (\psi^\dagger \partial_t \psi - (\partial_t \psi^\dagger)\psi) - {\cal H}_{\rm red}\right]
\ee
with
\be
{\cal H}_{red} (x,t) : = \frac{\hbar^2}{2} |\partial_x \psi |^2 + \frac{\lambda}{2} \pi^2 (x,t) + v J (x,t)
\ee
where
\be 
J(x,t) : = - \frac{i\hbar}{2} (\psi^\dagger \partial_x \psi - (\partial_x \psi^\dagger)\psi)
\ee
and
\be
\pi (x,t) : = - \frac{1}{2} \int dy \epsilon (x-y) J (y,t)
\ee
$\psi(x,t)$ has to be understood as a classical field -- because of the appearance of $\hbar$ in (5.10) we call it 'pseudo-classical'.

\medskip
\noindent
By varying the action $S$ with respect to $\psi^\dagger$ we obtain the following nonlinear, nonlocal Schr\"odinger equation for $\psi(x,t)$
\be
i\hbar (\partial_t - e(x,t) \partial_x - \frac{1}{2} (\partial_x e(x,t))) \psi (x,t) = - \frac{\hbar^2}{2} \partial^2_x \psi (x,t)
\ee
where in analogy to (4.11) we have
\be
e(x,t) : = - v + \frac{\lambda}{2} \int dy |x-y | J (y,t)\ .
\ee
On the other hand the variation of $S$ with respect to $v$ leads to the constraint
\be
\int dy J (y,t) = 0\ .
\ee

\noindent
Symplectic Hamiltonian analysis leads to the PB
\be
i\hbar \{ \psi (x,t), \psi^\dagger (y,t)\} = \delta (x-y)
\ee
with all other PB's vanishing.

\subsection*{C) Quantization and the $N$-particle amplitude}

The quantization of the Schr\"odinger equation (5.14) is obtained by means of the following canonical recipe:

\begin{description}
\item{i)} Substitute
\be
\left[ \psi (x,t), \psi^\dagger (y,t)\right] = \delta (x-y)
\ee
for the PB (5.17). All other commutators vanish.

\item{ii)} Keep all operator products in Wick-ordered form (i.e. shift all annihilation operators $\psi$ to the right)

\item{iii)} Consider the constraint (5.16) as a subsidary condition for the physical Hilbert space ${\cal G}$
\be
\int dy J (y,t) \phi = 0 \qquad \forall \phi \in {\cal G}
\ee
\end{description}
i.e. $\phi$ has vanishing total particle-momentum. 

The $N$-particle wave function is defined as usual by
\be
\chi_{_N} (x_1, \ldots, x_N,t) : = \frac{1}{\sqrt{N}} < 0 | \psi (x_1, t) \ldots \psi (x_N,t)| N >\ .
\ee  
The subsidary condition (5.19) applied to $|N>$ is then equivalent to
\be
\sum^N_{i=1} \partial_i \chi_{_N} = 0
\ee
i.e. $\chi_{_N}$ is only a function of the relative coordinates. This agrees with the classical situation described by (4.15-17).

Finally by means of (5.14), (5.18) and (5.21) we obtain the following Schr\"odinger equation for $\chi_{_N}$ as a function of relative coordinates $\{ \eta_i \}_{i=2}^N$
\begin{eqnarray}
i\hbar \partial_t \chi_{_N} &=& \left( - \frac{\hbar^2}{2} \partial_\alpha g^{\alpha\beta} (\eta) \partial_\beta - \frac{\hbar^2\lambda}{4} \sum_\alpha \delta (\eta_\alpha) - \right. \nonumber\\
& & \left. - \frac{\hbar^2\lambda}{4} \sum_{\alpha < \beta} \delta(\eta_\alpha - \eta_\beta)\right) \chi_{_N}\ .
\end{eqnarray}
This result corresponds to the following quantization rule for the classical $N$-particle Hamiltonian (4.16)
\be
{\rm H}_{class} = \frac{1}{2} g^{\alpha\beta} (\eta) \xi_\alpha \xi_\beta
\ee
\centerline{$\Downarrow$}
\begin{eqnarray}
{\rm H}_{quant} &=& - \frac{\hbar^2}{8} (\partial_\alpha \partial_\beta g^{\alpha\beta} (\eta) + g^{\alpha\beta} (\eta) \partial_\alpha \partial_\beta +\nonumber\\
&& + 2 \partial_\alpha g^{\alpha\beta} (\eta) \partial_\beta) \ .
\end{eqnarray}
The agreement of (5.24) with (5.22) is easily seen by means of the relation
\be
\partial_\alpha\partial_\beta (g^{\alpha\beta} (\eta)) = 2 \lambda (\sum_\alpha \delta(\eta_\alpha) + \sum_{\alpha < \beta} \delta (\eta_\alpha - \eta_\beta))\ .
\ee
Therefore in the case of the two-particle problem we have to choose $\gamma = \frac{1}{2}$ for the free parameter in (5.6). The quantization rule (5.23) $\Longrightarrow$ (5.24) is nothing but a generalization of this $\gamma = 1/2$ rule to a system of $N$ identical particles.

\bigskip
\noindent
Now we want to express our Hamiltonian operator as much as possible in terms of entities which are invariant with respect to arbitrary coordinate transformations on our (Pseudo-) Riemannian space $M$. This will be done by means of the following steps:

\begin{description}
\item{i)} Define $\varphi_{_N}$ by $\chi_{_N} = g^{1/4} \varphi_{_N}$. Then $\varphi_{_N}$ is normalized with respect to the invariant measure $\sqrt{g}$ $d\eta^1 \ldots d\eta^{N-1}$ 
\[
(g : = \det (g_{\alpha\beta}))
\]

\item{ii)} Decompose $M$ into $N$! regions specified by the signs of the $x_{ik}$. Then the $\delta$-terms in (5.22) give the corresponding boundary conditions for $\varphi_{_N}$ at the border between these regions.  On each of these regions our Hamilton operator takes the form
\be
H = -\frac{\hbar^2}{2} \Delta + \frac{\hbar^2}{8} R + V
\ee
where $\Delta$ is the invariant Laplace-Beltrami operator
\be
\Delta = \frac{1}{\sqrt{g}} \partial_\alpha g^{\alpha\beta} (\eta) \sqrt{g} \partial_\beta\ .
\ee
$R$ is the invariant scalar curvature (contraction of the Ricci-tensor) and $V$ is an additional noninvariant quantum correction potential defined in terms of Cristoffel symbols
\
\be
V = - \frac{\hbar^2}{8} \Gamma^\alpha_{\beta\mu} g^{\mu v} \Gamma^\beta_{\nu\alpha}\ .
\ee
Let us remark that the decomposition of the potential term in (5.26) into a part proportional to $R$ and a noninvariant residual $V$ is not unique. We defined this decomposition such, that $V$ is as simple as possible and the factor in front of $R$ agrees with a choice supported recently by De Witt [11]. 
\end{description}

\sec
\section{Two-particle bound states}

In this section we consider the bound state solutions of the stationary Schr\"odinger equation (5.22) for $N=2$
\be
E \chi_2 (x) = \{ - \hbar^2 \partial_x (1 + \frac{\lambda}{2} |x|) \partial_x - \frac{\hbar^2\lambda}{4} \delta (x) \} \chi_2 (x)\ .
\ee
In this case $M$ is the line which decomposes into the negative and positive half axis. With Bose-symmetry $\chi_2 (x) = \chi_2 (-x)$ we obtain from (6.1) on ${\bf R}^1_+$ the differential equation
\be
E \chi_2 = - \hbar \partial_x (1 + \frac{\lambda}{2} x) \partial_x \chi_2
\ee
with the boundary condition
\be
\partial_x \chi_2 (0) = - \frac{\lambda}{8} \chi_2 (0)\ .
\ee
The solutions of (6.2-3) may be smoothly continued to the negative real axis. Then the boundary condition (6.3) tells us, that the two particles obey fractional statistics (cp. [12]).

Let us now apply the procedure described in section 5.C:

\smallskip
\noindent
$M$ is flat and therefore $R = 0$. The Laplace-Beltrami operator is given by
\be
\Delta = - \hbar^2 (1 + \frac{\lambda}{2} x)^{1/2} \partial_x (1 + \frac{\lambda}{2} x )^{1/2} \partial_x
\ee
and we obtain for $V$
\be
V = - \frac{\hbar^2\lambda^2}{64}\ \frac{1}{1 + \frac{\lambda}{2} x}\ .
\ee
The boundary condition for $\varphi_2 (x) : = (2 + \lambda x)^{1/4}\chi_2$ at $x=0$ follows from (6.3) as 
\be
\partial_x \varphi_2 (0) = 0\ .
\ee
Because $M$ is flat we may again perform the transformation (4.40) to Euclidean space $E_1$ as in the classical case. This transforms our Schr\"odinger equation into its normal form
\be
E\tilde{\varphi}_2 (y) = (- \hbar^2 \partial^2_y + V(y) ) \tilde{\varphi}_2 (y)
\ee
with
\be
V(y) = - \frac{\hbar^2}{4y^2}
\ee
and the boundary condition
\be
\partial_y \tilde{\varphi}_2 (4/|\lambda|) = 0
\ee
The inner product has now to be taken with respect to the measure $dy$.

\medskip
In contrast to the classical case we have a nonzero potential in (6.7) which is a quantum correction. This term leads for scattering states to free particle behaviour with generalized statistics (cp. [13]). 

In order to proceed we have to treat the two cases $\lambda {>\atop <} 0$ separately.

\subsection*{A) $\lambda > 0$}

The solution of (6.7) which vanishes for $y \to \infty$ $(E < 0)$ is given by
\be
\tilde{\varphi}_2 (y) = y^{1/2} K_0 \left( \frac{\sqrt{-E}}{\hbar} y \right)
\ee
with the boundary condition (eq. (6.9))
\be
2 s K_1 (s) = K_0 (s)
\ee
where $s:= \frac{4}{\hbar\lambda} \sqrt{-E}$ and $K_{0,1}$ are modified Bessel functions of the third kind.

\medskip
\noindent
(6.11) has one fixed point only at
\[
s_0 = 0,16572
\]
i.e. there exists one bound state only.
This result corresponds to the well known fact that the $1d$-Schr\"odinger equation with a single $\delta$-potential has one bound state only. As the coupling strength of the $\delta$-potential in (6.1) is proportional to $\hbar^2$, the appearance of this bound state is a quantum effect. In the classical case only unbounded solutions with $E> 0$ are possible (cp. section 4.C).

\subsection*{B) $\lambda < 0$}

The requirement of finiteness of $\chi_{_2} (2/|\lambda|)$ leads to
\be
\tilde{\varphi}_2 (0) = 0
\ee
We have to consider the Schr\"odinger eq. (6.7) on the finite intervall $(0, 4/|\lambda |)$, because
the transformation (4.40) holds for $x \in I : = (0, 2/|\lambda|)$ only. From the physical point of view the singularity of the metric at $x = 2/|\lambda|$ means that  no events inside and outside of $I$ are connected to each other. In addition we remark that outside of $I$ our $H$ is not bounded from below and therefore unphysical. 

\medskip
\noindent
The solutions of (6.7) respecting the boundary conditions are $(E > 0)$
\[
\tilde{\varphi}_2 (y) = y^{1/2} J_0 \left( \frac{\sqrt{E}}{\hbar} y \right)
\]
with the spectrum given by the infinite number of fixed points of 
\be
2s J_1 (s) = J_0 (s)
\ee
with $s : = \frac{4}{\hbar(\lambda)} \sqrt{E}$. We obtain
\[
s_0 = 0,9407705640
\]
and
\be
s_n = \pi (\kappa_n + \frac{1}{4})
\ee
with 
\be
\kappa_n \stackrel{n\to\infty}{\longrightarrow} n\ .
\ee
Already for small values of $n$ the $\kappa_n$ lie close to the asymptotic values (6.15). We obtain
\begin{eqnarray}
\kappa_1 &=& 1,010306991\nonumber\\
&\vdots& \nonumber\\
\kappa_9 &=& 9.001369775 \qquad \mbox{etc.}
\end{eqnarray}
This infinite tower of bound states with $E>0$ corresponds to the classical situation which allows for bounded motion only (cp. eq. (4.39)). At the same time this tower gives us the quantum picture for our 'strings' of finite length with masses at their end points (cp. section 4.C).

\medskip
\noindent
This result also presents us with a quantum picture for our geometric bag model.

\sec
\section{Conclusions and outlook}

We have shown that the application of the gauge principle to $1d$-space translations for nonrelativistic point particles or fields leads to a nontrivial interaction with interesting features in both classical and quantum dynamics. But to which interaction known in nature corresponds this interaction? We don't know of any macroscopic interaction between two particles with a potential proportional to their distance and energy as in (4.38). For a gauge coupling $\lambda > 0$ this interaction is unphysical anyway because it leads to arbitrary large velocities if time goes on. The same holds for $\lambda < 0$ if the interparticle distance is larger then $2/|\lambda|$. On the other hand we have shown in section 4.D that classically we obtain for $\lambda < 0$ a geometric bag for two and three particles respectively. This result has been quantum mechanically extended for $N=2$ in section 6. Therefore we may speculate, that our new interaction is of physical relevance for $\lambda < 0$ in the microscopic regime in connection with the confinement problem. But such a speculative statement immediately leads to two new and important questions:

\begin{description}
\item{i)} What happens in higher space dimensions $d=2$ or $3$\ ? Will geometric bag formation persist?

Work for $d=2$ with a Chern-Simons-like interaction for the gauge fields is in progress. Thereby we will also include the second central charge of the Galilei group in the free Lagrangian (cp. [14]). In this context we will also take up the problem of particles and fields moving on a circle again.

\item{ii)} What is the connection with QCD? For that we have to extend our framework by supplying our particles and fields respectively with a nonabelian color charge and by gauging the corresponding $SU(N)$-group simultaneously with the space translations.

We will take up this very interesting question as soon as possible. 
\end{description}

\bigskip
\medskip
\noindent
{\bf Acknowledgement}

\medskip
\noindent
I'm very grateful to J.~Lukierski for extensive discussions and valuable hints and to E.H.~de Groot for helpful comments.

\bigskip
\bigskip
\noindent
{\Large\bf References}

\medskip
\noindent
\begin{description}
\item{[1]} R.~Mills, Am.~J.~Phys.~{\bf 57} (1989) 493,\\
F.~Gronwald and F.W.~Hehl, in ``Quantum gravity" (P.G.~Bergmann, V.~de Sabbata and H.J.~Treder Eds.), World Scientific, Singapore, 1996.

\item{[2]} R.~De Pietri, L.~Lusanna and M.~Pausi, Class.~Quant.~Grav.~{\bf 12} (1995), 219 and 255; {\bf 13} (1996), 1417.

\item{[3]} P.C.~Stichel, Phys.~Lett.~{\bf B456} (1999), 129 (hep-th/9902207).

\item{[4]} D.~Cangemi, R.~Jackiw and B.~Zwiebach, Ann.~Phys.~(N.Y.) {\bf 245} (1996), 408.\\
R.B.~Mann and T.~Ohta, Class.~Quant.~Grav.~{\bf 14} (1997), 1259.

\item{[5]} D.~Cangemi and R.~Jackiw, Ann.~Phys.~(N.Y.) {\bf 225} (1993), 229.

\item{[6]} R.~Jackiw and A.P.~Polychronakos, in ``Faddeev Festschrift", Steklov Mathematical Institute Proceedings, preprint hep-th/9809123\\
N.~Ogawa, preprint hep-th/9801115.

\item{[7]} Yu.N.~Obukhov and S.N.~Solodukhin, Class.~Quant.~Grav.~{\bf 7} (1990), 2045. 

\item{[8]} S.~Silva, Nucl.~Phys.~{\bf B558} (1999), 391.

\item{[9]} L.D.~Faddeev and R.~Jackiw, Phys.~Rev.~Lett.~{\bf 60} (1988), 1692.

\item{[10]} E.~Madelung, Z.~Phys.~{\bf 40} (1926), 322.

\item{[11]} B.S.~De Witt, Supermanifolds, 2nd ed., Cambridge University Press, Cambridge, 1992.

\item{[12]} J.M.~Leinaas and J.~Myrheim, Int.~J.~Mod.~Phys.~{\bf B5} (1991), 2573.

\item{[13]} A.P.~Polychronakos, Generalized statistics in one dimension, Les Houches Lectures, Summer 1998, preprint hep-th/9902157. 

\item{[14]} J.~Lukierski, P.C.~Stichel and W.~Zakrzewski, Ann.~Phys.~(N.Y.) {\bf 260} (1997), 224.
\end{description}

\end{document}